\documentclass[
    ,final            
  ]
  {aipproc}

\layoutstyle{6x9}

\begin{document}

\title{
 Measurement of Double Helicity Asymmetry in Multi-Particle
 Production with Polarized Proton-Proton Collision at PHENIX
}

\classification{
  14.20.Dh, 13.88+e, 13.85.Hd
}
\keywords      {
  Proton Spin Structure, Polarized Gluon Distribution Function
}

\author{Kenichi Nakano on behalf of the PHENIX Collaboration}{
  address={Tokyo Institute of Technology, 2-12-1 Ookayama, Meguro-ku,
  Tokyo 152-8550, Japan}
}

\begin{abstract}
 A goal of the PHENIX experiment is to obtain the polarized
 gluon distribution function in the proton.
 Double helicity asymmetry in multi-particle production 
 with polarized proton-proton collision is measured at
 midrapidity with RHIC Run 2005 data.
 This result excludes the maximum positive gluon polarization
 (``GRSV-max'').
\end{abstract}

\maketitle


\section{INTRODUCTION}

Polarized deep inelastic lepton-hadron scattering experiments revealed
that the contribution of the quark spin to the proton spin is only
20-30\% \cite{Ashman:1987hv}\cite{:2006vy}.
The remaining component can be carried by the gluon spin and the angular
momenta of quarks and gluons.
One of the goals of the PHENIX experiment is to obtain the contribution
of the gluon spin, namely the polarized gluon distribution function in
the proton, $\Delta g(x)$.
$\Delta g(x)$ is evaluated by measuring double helicity asymmetry,
$A_{LL}$, of reactions in longitudinally-polarized proton-proton
collisions, for example, jet, $\pi^0$ or direct photon productions.
A jet is measured as a cluster of multiple particles.
The particle cluster measurement gives higher statistics at high
transverse momentum ($p_T$) than single particle measurements such as
$\pi^0$.
$A_{LL}$ is defined as
\begin{equation}
A_{LL} = \frac{1}{|P_B| |P_Y|} \frac{N_{++} - R N_{+-}}{N_{++} + R N_{+-}}.
\end{equation}
Here, $N_{++}$ and $N_{+-}$ are the number of measured particle
clusters with same and opposite beam helicities, respectively;
$R \equiv L_{++}/L_{+-}$ is relative luminosity;
and $P_B$ and $P_Y$ are beam polarizations.

\section{EXPERIMENTAL SETUP}

We used proton-proton collision data taken in 2005 at 
$\sqrt{s} = 200$ GeV with an average beam polarization of $\sim$46\%.
The integrated luminosity of analyzed data is 2.2 pb$^{-1}$.
The PHENIX Central Arm detectors were used.
Two arms are positioned almost back to back, 
and each arm covers the pseudorapidity region 
$|\eta| < 0.35$ and the 90-degree azimuthal angle.
Photons with $p_T > 0.4$ GeV/$c$ were measured with electromagnetic
calorimeters, and an electromagnetic-shower-shape cut and an
charged-track-matching veto were applied to eliminate clusters made by
charged particles.
Charged particles with $0.4 < p_T < 4$ GeV/$c$ were measured with drift
chambers and pad chambers.
For each event a high-$p_T$ ($>2$ GeV/$c$) photon was required to exist 
so that the efficiency of high-$p_T$-photon trigger becomes $p_T$
independent.

\section{METHODS OF PARTICLE CLUSTER MEASUREMENT}

Particles that satisfied the experimental selections were clustered
by a cone method.
Starting from each particle in an event as a seed,
an interative procedure was used to define a cone around
a cluster of particles.
The cone radius $R$ ($= \sqrt{(\Delta \phi)^2 + (\Delta \eta)^2}$)
was set to 0.3.
The transverse momentum of the cone, $p_T^{\rm cone}$, was defined as
the vector sum of the transverse momenta of the particles in the cone;
\begin{equation}
p_T^{{\rm cone}} \equiv \left| \sum_{i \in {\rm cone}} \vec{p}_{Ti} \right|
\end{equation}
The cone that gives the largest $p_T^{\rm cone}$ in an event was used
in the event.

The relationship between $p_T^{\rm cone}$ and $p_T^{\rm jet}$ was
evaluated with PYTHIA and GEANT simulations \cite{Nakano:2005jps}.
From two hard-scattered partons in PYTHIA jet event we selected one
parton closer in $R$ to cone axis, and adopted $p_T$ of the parton as
$p_T$ of jet.
We used PYTHIA version 6.220 with the ``Rick Field MPI tune A'' 
setting \cite{Field:2005qt}.
We have confirmed that the simulation reproduced the event shape,
such as particle multiplicity, one-sided thrust in the PHENIX Central
Arm acceptance and particle $p_T$ density, 
reasonably well \cite{Nakano:2005jps}.

The largest systematic uncertainty in this measurement is an uncertainty
on the $p_T$ scale difference between measurement and theory.
It originates in the fact that a jet is defined with cone at hadron
level in measurement and at parton level in theory.
In $A_{LL}$ measurement we assigned the $p_T$ scale uncertainty 10\%, by
which the $p_T$ scale in theory varies when a theory cone size is
changed in a typical range.
We could checked the $p_T$ scale uncertainty by measuring cross section
or a similar quantity because the difference in $p_T$ scale appears as a
shift of cross section curve.

The requirement of a high-$p_T$ photon causes a bias on the fractions of
jet production subprocesses (quark-quark, quark-gluon and gluon-gluon
reactions).
This effect was estimated using PYTHIA.
In general, the gluon-gluon reaction is suppressed particularly at
low $p_T$.
No systematic uncertainty on it has been assigned because this effect is
much smaller than other systematic uncertainties.

\section{RESULT}

\begin{figure}
\includegraphics[width=0.8\linewidth]{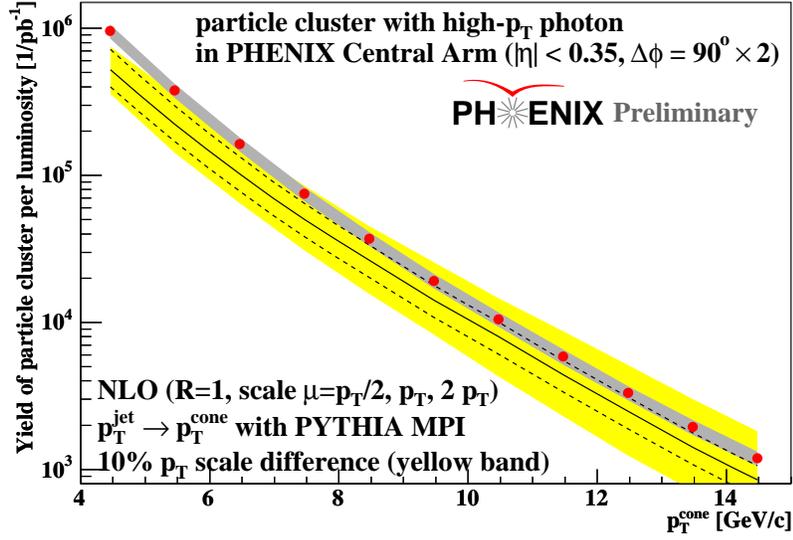}
\caption{
 Yield per luminosity of particle clusters.
 {\bf Red points}: real data with systematic error as gray band.
 {\bf Black lines}: predicted ones evaluated with the NLO jet cross
 section  \cite{Jager:2004jh} and the $p_T^{\rm jet}$-$p_T^{\rm cone}$
 relationship.
 {\bf Yellow band}: deviation of solid line by $\pm$10\% $p_T$ scale
 variation.
}
 \label{fig:evtrate_multiparticle}
\end{figure}

Figure \ref{fig:evtrate_multiparticle} shows yield per luminosity of
particle cluster with high-$p_T$ photon trigger in the PHENIX Central
Arm acceptance as a function of $p_T^{\rm cone}$.
The real data is drawn as red points with systematic errors as a gray
band.
The main systematic errors on real data are luminosity uncertainty
(10\%) and EMCal energy scale uncertainty (5\%).
The predicted ones are drawn as three black lines,
which are based on the NLO jet cross section with $R = 1$ at 
$\left| \eta \right| < 0.35$ \cite{Jager:2004jh} and the
$p_T^{\rm jet}$-$p_T^{\rm cone}$ relationship.
The yellow band shows the deviation of the black solid line caused by a
$\pm$10\% $p_T$ scale variation.
The difference in the yield per luminosity between the real data and the
predicted one doesn't exceed the deviation.

\begin{figure}
\includegraphics[width=0.8\linewidth]{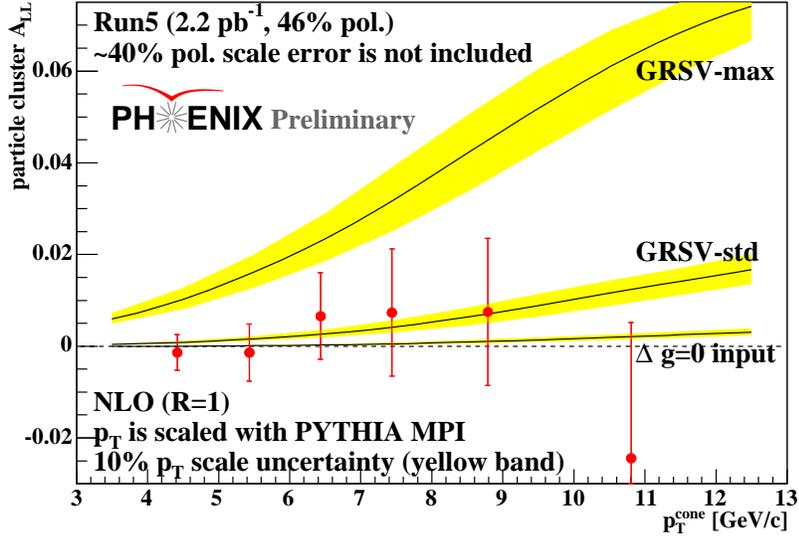}
\caption{
 Particle cluster $A_{LL}$.
 {\bf Red points}: real data.
 {\bf Black lines}: predicted ones evaluated with the NLO jet $A_{LL}$
 \cite{Jager:2004jh}, the three $\Delta g(x)$ models \cite{Gluck:2000dy}
 and the $p_T^{\rm jet}$-$p_T^{\rm cone}$ relationship.
 {\bf Yellow band}: systematic uncertainty on $p_T$ scale difference
 between measurement and theory.
}
 \label{fig:asym_multiparticle}
\end{figure}

Figure \ref{fig:asym_multiparticle} shows the $A_{LL}$ of particle
cluster production as a function of $p_T^{\rm cone}$.
The predicted $A_{LL}$'s are drawn as black lines, 
which are based on the NLO jet $A_{LL}$ with $R = 1$ at 
$\left| \eta \right| < 0.35$ \cite{Jager:2004jh}, the
three $\Delta g(x)$ models \cite{Gluck:2000dy} and 
the $p_T^{\rm jet}$-$p_T^{\rm cone}$ relationship.
The yellow band shows the 10\% $p_T$ scale uncertainty.
This result excluded the ``GRSV-max'' case and indicated a similar
probability for the ``GRSV-std'' and ``$\Delta g = 0$ input'' cases.

\section{CONCLUSIONS}

Double helicity asymmetry in multi-particle production was measured at
midrapidity reagion ($\eta < 0.35$) at $\sqrt{s} = 200$ GeV.
The analyzed data were taken in 2005 with $L = 2.2$ pb$^{-1}$ and
$\bar{P} \sim$ 46\%.
The relationship between $p_T^{\rm jet}$ and $p_T^{\rm cone}$ was
evaluated with the PYTHIA and GEANT simulations.
The systematic uncertainty on $A_{LL}$ was assigned 10\% due to the $p_T$
scale difference between measurement and theory.
This result excluded the ``GRSV-max'' case and indicated a similar
probability for the ``GRSV-std'' and ``$\Delta g = 0$ input'' cases.
It will make stronger constraint on $\Delta g(x)$ with futher
statistics.


\begin{theacknowledgments}
Data analyses and simulation studies have been done at RIKEN CCJ.
\end{theacknowledgments}

\bibliographystyle{aipproc}   

\begin{thebibliography}{9}
\bibitem{Ashman:1987hv}
J.~Ashman {\it et al.}  [European Muon Collaboration],
Phys.\ Lett.\ B {\bf 206}, 364 (1988).

\bibitem{:2006vy}
[HERMES Collaboration],
arXiv:hep-ex/0609039.



\bibitem{Nakano:2005jps}
K.~Nakano,
Talk at Second Joint Meeting of Nuclear Physics Divisions of APS and
JPS, Session BG, 2005.

\bibitem{Field:2005qt}
R.~Field  [CDF Collaboration],
Acta Phys.\ Polon.\ B {\bf 36}, 167 (2005).


\bibitem{Jager:2004jh}
B.~Jager, M.~Stratmann and W.~Vogelsang,
Phys.\ Rev.\ D {\bf 70}, 034010 (2004)
[arXiv:hep-ph/0404057].

\bibitem{Gluck:2000dy}
M.~Gluck, E.~Reya, M.~Stratmann and W.~Vogelsang,
Phys.\ Rev.\ D {\bf 63}, 094005 (2001)
\end{thebibliography}

\end{document}